\documentclass[onecolumn,a4paper,12 pt,doublespace]{article}

\usepackage{graphicx}
\usepackage{wrapfig}

\begin{document}

\begin{center}
\textbf {Efficient collinear third-harmonic generation in a single two-dimensional nonlinear photonic crystal}
\end{center}

\begin{center}
\footnotesize Todor Karaulanov, Solomon M. Saltiel\\
Faculty of Physics, University of Sofia, 5 J. Bourchier Boulevard, Sofia 1164, Bulgaria,\\Tel +359-2 62 15 30 e-mails: tod\_kulan@hotmail.com, saltiel@phys.uni-sofia.bg\end{center}

\vspace {3 cm}

\begin{abstract}
We propose novel multi-phase-matched process that starts with generation of a pair of symmetric second-harmonic waves.  Each of them interacts again with the fundamental wave to produce two constructively interfering third harmonic waves collinear to the fundamental input wave. 
\end{abstract}
\newpage
\begin{center}
\textbf {Efficient collinear third-harmonic generation in a single two-dimensional nonlinear photonic crystal}
\end{center}

\begin{center}
\footnotesize Todor Karaulanov, Solomon M. Saltiel\\
Faculty of Physics, University of Sofia, 5 J. Bourchier Boulevard, Sofia 1164, Bulgaria,\\\end{center}

 Nonlinear photonic crystals have attracted much interest lately because of their unique properties and many potential applications in nonlinear optics and  optics communications. Two dimensional nonlinear photonic crystals (2DNPC) [1] allow realization of simultaneous phase matching of several processes [2], however in most of the cases generated wave is noncollinear to the input fundamental wave that was considered as a disadvantage for 2DNPC. Here we propose for the first time phase-matching geometry for third harmonic (TH) generation in 2DNPC, for which generated TH wave is collinear to the input wave. In addition to that the proposed interaction at low input intensities is four times more efficient than the conventional schemes.

\begin{wrapfigure}{r}[0pt]{0.5\textwidth}
   \centerline{\includegraphics[width=0.5\textwidth]{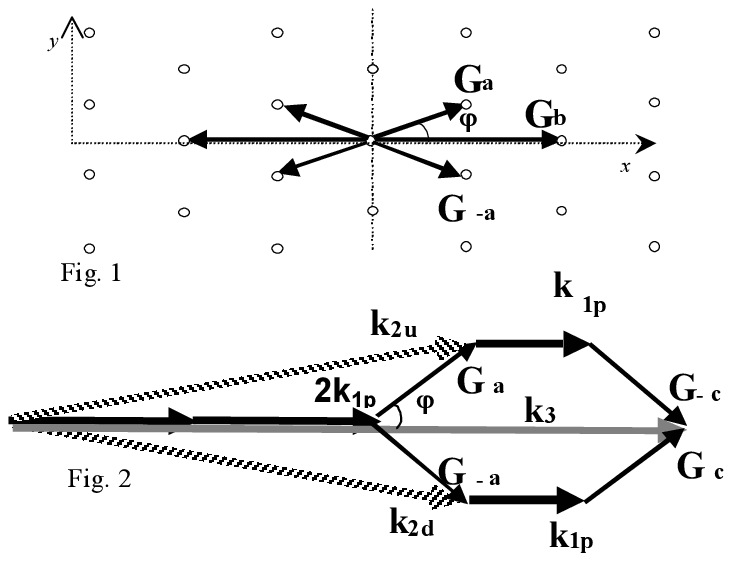}}
   \vspace*{3mm}~
\end{wrapfigure}

Reciprocal lattice of the 2DNPC  with  the main reciprocal wave-vectors $G_a, G_b, G_{-a}$  are shown on the Fig. 1. Multiple phase matching geometry is presented on Fig. 2. The process starts with generation of pair second harmonic waves with wave vectors  $k_{2u}, k_{2d}$. Phase matching is assured by the reciprocal fundamental vectors  $G_a, G_{-a}$ . Each second harmonic wave interacts again with the fundamental wave via pair phase matched interaction with participation of the reciprocal vectors   generating this way pair collinear TH waves with one and the same wave vector $k_{3}$.  The reciprocal vectors that phase match the processes of sum frequency mixing $\omega + 2\omega = 3 \omega$  are combination of the two fundamental reciprocal vectors $G_c=NG_b-G_{a}$ ; $G_{-c}=NG_b-G_{-a}$. The two TH waves interfere constructively in the direction of the input fundamental wave leading this way to higher overall TH efficiency. 
The process is described by system of 5 differential equations, that have been solved in several approximation. Design and optimization of TH frequency converters made from 2DNPC implemented in LiNbO$_3$ and LiTaO$_3$ is presented.
The frequency converter can be also useful for testing the homogeneity of the sample. The better is the homogeneity the higher will be TH efficiency.\\*[12pt]
\footnotesize 1. V. Berger, Phys. Rev. Lett. 81, 4136 (1998).\\
2. S. Saltiel and Yu. S. Kivshar, Opt. Lett. 25, 1204 (2000).

\end{document}